\documentclass[12pt]{article}

\usepackage{amssymb,amsmath,xcolor,graphicx,xspace,colortbl, rotating} 
\usepackage{amsfonts}  
\usepackage{amsmath}  
\usepackage{amssymb}  
\usepackage{epsfig,graphicx}  
\usepackage[authoryear]{natbib}
\usepackage{latexsym}  
\usepackage{rotating}  
\usepackage{setspace}  

\graphicspath{{MMLV_graphics/}{MMLV_tcache/}{MMLV_gcache/}}
\DeclareGraphicsExtensions{.pdf,.svg,.eps,.ps,.png,.jpg,.jpeg}
\usepackage {bm}
\usepackage {tabulary}
\newtheorem {condition}{Assumption}

\newtheorem {definition}{Definition}

\newtheorem {lemma}{Lemma}

\newtheorem {proposition}{Proposition}

\begin{document}

\title{Identification of Unobservables in Observations}

\author{Yingyao\ Hu\protect\footnote{Contact information: Department of Economics, Johns Hopkins University,
3400 N. Charles Street, Baltimore, MD 21218. Email: yhu@jhu.edu.  } ~\\
Johns Hopkins University}
\date{\today}
\maketitle
\begin {abstract}

In empirical studies, the data usually don't include all the variables of interest in an economic model. This paper shows the identification of unobserved variables in observations at the population level. When the observables are distinct in each observation, there exists a function mapping from the observables to the unobservables. Such a function guarantees the uniqueness of the latent value in each observation. The key lies in the identification of the joint distribution of observables and unobservables from the distribution of observables. The joint distribution of observables and unobservables then reveal the latent value in each observation. Three examples of this result are discussed.

\bigskip JEL classification:\textit{\ C01, C14}

Keywords:\textit{ identification in observations,  unobserved heterogeneity , latent variable model, measurement
error model}

\end {abstract}

\setcounter{page}{0}
\thispagestyle{empty}

\newpage
\setstretch {1.15}

\section{Introduction}

\begin{center}
\textit{``No two leaves are alike."}
\end{center}

Thousand years of human history show that no two leaves are alike.  Suppose that each leaf has observed traits $x$ and unobserved heterogeneity $x^*$. If we observe that no two leaves are alike, then the mapping from $x$ to $x^*$ is a function, i.e., a set of ordered pairs $(x,x^*)$ in which no two different ordered pairs have the same first coordinate. Therefore, such a function can uniquely determine unobserved heterogeneity $x^*$ from observed traits $x$ for each leaf.

In empirical studies, an observation in the data is similar to observed traits $x$ of a leaf and unobserved heterogeneity $x^*$ corresponds to some variables of interest not observed in the data. For example, researchers observe a patient's insurance policy, but not their health status in the data. In general, we observe an agent's choices, but not their types or unobserved heterogeneity. In macroeconomics, we are interested in a country's true GDP when only rough measurements are available. This paper intends to provide a framework to identify the value of a latent variable of interest in observations. 

For a variable with a distinct value in each observation in a sample, researchers usually consider it as a continuous variable in the population. Such continuity only exists in assumptions given the discrete nature of a sample. It is observationally equivalent to assume that the population is a collection of a large but finite number of elements. To avoid an uncountable amount of unknowns, we adopt the latter in this paper.  

Let $x_i  $ and $x^*_i $ be measurements of observed traits and unobserved heterogeneity of leaf $i$, respectively. We define the property of leaves as follows:
\begin{definition} \label{def leaves}
A population $\mathcal{P}_{X,X^*}$ satisfies \textbf{the property of leaves} if it  is a collection of ordered pairs $(x_i, x^*_i)   $ for $i= 1,2,..., N; N < \infty $ such that  $x_i \neq x_j$  for any $i \neq j$. That is
\begin{equation} \label{definition 1}
\mathcal{P}_{X,X^*} = \{(x_i, x^*_i) :  x_i \neq x_j  \, for \, i \neq j \,and \, i,j= 1,2,..., N. \}
\end{equation}
\end{definition}
Furthermore, let $F_{X,X^*}$ denote the cumulative distribution function of random variables $(X, X^*)$ randomly drawn from population $\mathcal{P}_{X,X^*}$ with probability 
\begin{equation} \label{probability p}
Pr(\{(X,X^*)=(x_i,x^*_i)\})=p_i>0
\end{equation}
 with $\sum_{i=1}^{N} p_i = 1$. A typical example is $p_i = \frac{1}{N}$. Here we focus on the case where the population size $N$ is large but finite. In this case, distribution function $F_{X,X^*}$ uniquely determines population $\mathcal{P}_{X,X^*}$ because $F_{X,X^*}$ is a step function and each step corresponds to an element in $\mathcal{P}_{X,X^*}$. If we consider the set $\mathcal{P}_{X,X^*}$ as a mapping from $X$ to $X^*$, then this mapping is a function, i.e., a set of ordered pairs in which no two different ordered pairs have the same first coordinate. The population of observed traits $x$ is
\begin{equation}  \label{definition 2}
\mathcal{P}_X = \{x_i :  (x_i, x_i^*) \in \mathcal{P}_{X,X^*} \, for \, some \, x_i^* \}
\end{equation}
with a distribution function $F_{X}$. In fact, its probability function is 
\begin{equation} 
Pr(\{X=x_i\})=p_i
\end{equation}
because $x_i$ is distinct for all $i=1,2,...,N$, i.e., in the whole population. Therefore, the probabilities $p_i$ is known for each $i$ from $F_{X}$, but the value of $x^*_i$ in Equation (\ref{probability p}) still needs to be identified and distribution function $F_{X,X^*}$ is still unknown. Notice that $x^*_i$ is not necessarily distinct in observations. 

Here it is necessary to clarify that this identification analysis is at the population level, instead of at the sample level. In estimation, we usually start with a sample of $X$ and use its sample statistics to estimate its population counterparts.\footnote{If we draw from such a population of $X$ with placement, it is possible to have two draws with the same value $x$. The property of leaves guarantees that the value $x$ corresponds to a unique $x^*$ and that the two draws are from the same leaf, which can be represented in the sample proportions of each distinct value of $x$. Therefore, such a generated sample of $X$ is representative of the population of leaves and its distribution.}  For example, we use empirical CDF $\hat{F}_{X}$ to consistently estimate the population CDF $F_{X}$. In the identification analysis, we start with population $\mathcal{P}_X$ and its CDF $F_{X}$ when $X$ is observed in a sample because $\mathcal{P}_X$ and $F_{X}$ are identified as the limit of the sample and the empirical CDF, respectively. This paper presents sufficient conditions, under which one can uniquely determine the unobserved $x^*$  from the observed $x$ in each observation in the population. We summarize the immediate conditions as follows:

\begin{proposition} \label{proposition 1}
Suppose that Conditions 1 and 2 hold as follows:
\begin{enumerate}
\item  Population $\mathcal{P}_{X,X^*}$, with distribution function  $F_{X,X^*}$,   satisfies the property of leaves in Equations  (\ref{definition 1}) ; 
\item   $F_{X}$ uniquely determines  $F_{X,X^*}$, where distribution function  $F_{X}$ corresponds to population $\mathcal{P}_{X}$  in Equations  (\ref{definition 2}).
\end{enumerate}
Then, $\mathcal{P}_{X}$ and $F_{X}$ uniquely determine $\mathcal{P}_{X,X^*}$ and $F_{X,X^*}$, i.e., each $x_i$ in $\mathcal{P}_{X}$ uniquely determines its corresponding $x^*_i$ through $\mathcal{P}_{X,X^*}$.
\end{proposition}

\textbf{Proof}: Distribution function $F_{X}$ corresponds to population $\mathcal{P}_{X}$. Condition 2) requires that $F_{X}$ uniquely determines  $F_{X,X^*}$. Given that population $\mathcal{P}_{X,X^*}$ contains a large but finite number $N$ of different elements, distribution function $F_{X,X^*}$ is a step function and each step corresponds  to one element in population $\mathcal{P}_{X,X^*}$, and therefore,  $F_{X,X^*}$ uniquely determines  population $\mathcal{P}_{X,X^*}$. In summary, $\mathcal{P}_{X}$ with  $F_{X}$ uniquely determines $\mathcal{P}_{X,X^*}$ with $F_{X,X^*}$. Q.E.D.

\bigskip

Although it is for a large but finite $N$, this result can be extended to the case with $N \rightarrow \infty$ as long as population $\mathcal{P}_{X,X^*}$ contains a countably number of elements. Note that the probability in Equation (\ref{probability p}) can't be uniform, i.e., $p_i=p$, in this case. The discreteness of $\mathcal{P}_{X,X^*}$ implies that population distribution $F_{X,X^*}$ is a step function and each step still corresponds one element in population $\mathcal{P}_{X,X^*}$. Such a property is lost when there are uncountably many elements in the population, e.g., the unit interval. \footnote{A recent working paper \cite{HuLiuYao2022} provides some identification arguments in that case.} 

Conditions 1) in Proposition \ref{proposition 1} requires that the observed $X$ should satisfy the property of leaves so that there exists a function mapping from the observed to the unobserved. Condition 2 is the key to achieve the identification of unobservables in observations. In order to make Proposition \ref{proposition 1} useful, it is important to  provide sufficient conditions to identify $F_{X,X^*}$ from $F_{X}$. 

Proposition \ref{proposition 1}  can be adapted to the case where additional variables are observed. If the data include $(X,Z)$ instead of $X$ only, and if $F_{X,Z}$ uniquely determines $F_{X,X^*}$, then there is no need to identify $F_{X,Z,X^*}$, which may require more assumptions than those for the identification of $F_{X,X^*}$. In that case, the result in Proposition \ref{proposition 1} remains with $F_{X,Z}$ uniquely determining $F_{X,X^*}$ in Condition 2). 

It is useful to understand the result in Proposition \ref{proposition 1} in the case of the widely-used linear regression model, i.e.,
$$
Y=W\beta + \eta
$$
with $E[\eta|W]=0$. A researcher observes $X=(Y,W)$ but not $X^*=\eta$. Suppose that we never observe repeated values of $(Y, W)$ in the population. Therefore, Condition 1) is satisfied. In fact, the function implied by Condition 1), which maps from $X=(Y,W)$ to $X^*=\eta$, is given by the model, i.e., $\eta=Y-W\beta$.  Given that we can identify and estimate $\beta$ using distribution $F_{Y,W}$ through the moment equation $E[Y-W\beta|W]=0$, parameter $\beta$ can be considered as known from the population. Then, it can be shown that $F_{X}$, i.e., the distribution of $(Y,W)$, uniquely determines  $F_{X,X^*}$, i.e., the distribution of $(Y,W, \eta)$, because $\eta=Y-W\beta$, and therefore, $\mathcal{P}_{X}$, the population of $(Y,W)$, uniquely determines $\mathcal{P}_{X,X^*}$, the population of $(Y,W,\eta)$. That means, the regression error $\eta_i $, although unobserved, is uniquely determined by $(y_i,w_i)$ in each observation through $\mathcal{P}_{X,X^*}$ as $\eta_i = y_i - w_i \beta$. The estimation of residuals in the linear regression model is the sample counterpart of this procedure.

Condition 1) in Proposition \ref{proposition 1} holds as long as there is a traditionally-defined continuous variable in the sample. The challenging part of Proposition \ref{proposition 1} is to show that the joint distribution of observables and unobservables is uniquely determined by that of the observables. The next two sections present examples of sufficient conditions for the identification of $F_{X,X^*}$ from $F_{X}$. We adopt the framework in \cite{hu2017econometrics}  to consider cases with a difference number of measurements of $X^*$ in $X$.

\section{A 2-measurement case}

In this section, we consider a 2-measurement setting, where 
$
X=(X_1,X_2).
$
Assume that $X_1$, $X_2$, and $X^*$ are a scalar random variable satisfying \begin{eqnarray}  \label{equ 150}
X_1 &  = & X^{ \ast } +\epsilon_1 \nonumber  \\
X_2 &  = & X^{ \ast } +\epsilon_2 
\end{eqnarray}  
where i) $\epsilon_1 $ is independent of $(X^{ \ast }, \epsilon_2)$, ii) the characteristic function of $X_1$ is absolutely integrable and does not vanish on the real line, and iii) $E[\epsilon_2| X^{ \ast }]=0$. 

Before presenting the technical results, it is useful to illustrate the idea of identification in observations with a simple example. Suppose $X^{ \ast } \in \{0,1\} $, $\epsilon_1 \in  \{-1,2\}$ and $\epsilon_2 \in  \{-1, 0,1\} $ satisfying Equation (\ref{equ 150}) with distribution functions $ f_{X^*, \epsilon_2}$ and $f_{\epsilon_1}$. Notice that $\epsilon_2 $ should have a zero mean conditional on $X^*$. The population and its distribution can be presented as in Table \ref{table 1}. Given the population of 12 observations of $(X_1,X_2)$ with distribution function $f_{X_1,X_2}$, the goal is to show that the value of $X^*$ is uniquely determined in each observation.

\begin{table}[htp]
\caption{An illustration of identification in observations}
\begin{center}
\begin{tabular}{c|cc|ccc|c}
\hline 
observation & \multicolumn{2}{c|}{observables} &\multicolumn{3}{c|}{unobservables}& probability\\
$i$ & $X_1=X^{ \ast }+\epsilon_1$ & $X_2=X^{ \ast }+\epsilon_2$  & $\epsilon_1$ & $X^{ \ast }$ &  $\epsilon_2$ & $p_i$ \\
\hline
1 & 0 & 0 & -1 & 1 & -1 & \tiny $ f_{X_1,X_2}(0,0)= f_{\epsilon_1}(-1)   f_{X^*,\epsilon_2}(1,-1)$ \normalsize \\
2 & 0 & 1 & -1 & 1 & 0 & \tiny  $f_{X_1,X_2}(0,1)= f_{\epsilon_1}(-1)   f_{X^*,\epsilon_2}(1,0)$ \normalsize \\
3 & 0 & 2 & -1 & 1 & 1 & ... \\
\hline
4 & -1 & -1 & -1 & 0 & -1 &  ... \\
5 & -1 & 0 & -1 & 0 & 0 &  ... \\
6 & -1 & 1 & -1 & 0 & 1 &  ... \\
\hline
7 & 3 & 0 & 2 & 1 & -1 &  ... \\
8& 3 & 1 & 2 & 1 & 0 &  ... \\
9 & 3 & 2 & 2 & 1 & 1 &  ... \\
\hline
10 & 2 & -1 & 2 & 0 & -1 &  ... \\
11 & 2 & 0 & 2 & 0 & 0 &  ... \\
12 & 2 & 1 & 2 & 0 & 1 &  ... \\
\hline
\end{tabular}
\\ Note: In each group, mean of $X_2$ reveals $X^*$.
\end{center}
\label{table 1}
\end{table}%

In this example, the observed $(X_1,X_2)$ are distinct. If we group the 12 observations by $X_1$, then the mean of $X_2$ within each group is equal to the value of latent $X^*$. In general, the property of leaves guarantees the uniqueness of $X^*$ in each observation. The restrictions on the distribution, i.e., $\epsilon_2 $ should have a zero mean conditional on $X^*$, reveal the value of $X^*$ in each observation. Therefore, the unobserved is uniquely determined by the observed in observations. Notice that the four groups are actually categorized by the values of $X^*$ and $\epsilon_1$. 

Although we use $X_2$ to put the 12 observations into 4 groups in this particular example in Table \ref{table 1}, it really is the combination of the 12 observations of $(X_1,X_2)$ and the distribution function of $f_{X_1,X_2, X^*}$ or $f_{\epsilon_1}f_{X^*,\epsilon_2}$ that identifies the 4 groups out of the 12 observations. It is possible that $X_1$ is not enough to distingish all the groups with the same values of $X^*$ and $\epsilon_1$. For example, we may change the support of $\epsilon_1$ and $\epsilon_2$ to be $\epsilon_1 \in  \{-1,0\}$ and $\epsilon_2 \in  \{-1.5,0.5,1\}$, still assuming $\epsilon_2 $ should have a zero mean. Table \ref{table 1b} shows the population and the probabilities in this case, where $X_1$ is no longer enough to identify the four groups. From the observed 12 probabilities, i.e., $p_i$, in $f_{X_1,X_2}$, however, we are able to identify the 8 unknown probabilities in  $f_{\epsilon_1}$ and $f_{X^*,\epsilon_2}$, which will be shown below in a more general setup.  The identified distribution function, i.e., the probabilities in  $f_{\epsilon_1}$ and $f_{X^*,\epsilon_2}$ can determine how to put the observations into four groups with the same $X^*$ and $\epsilon_1$ as in the last column in Table \ref{table 1b}. Then, mean of $X_2$ reveals $X^*$ in each group.

\begin{table}[htp]
\caption{A second example}
\begin{center}
\begin{tabular}{c|cc|ccc|c}
\hline 
observation & \multicolumn{2}{c|}{observables} &\multicolumn{3}{c|}{unobservables}& probability\\
$i$ & $X_1=X^{ \ast }+\epsilon_1$ & $X_2=X^{ \ast }+\epsilon_2$  & $\epsilon_1$ & $X^{ \ast }$ &  $\epsilon_2$ & $p_i$ \\
\hline
1 & 0 & -0.5 & -1 & 1 & -1.5 & \tiny  $ f_{X_1,X_2}(0,-0.5)= f_{\epsilon_1}(-1)  f_{X^*,\epsilon_2}(1,-1.5)  $ \normalsize \\
2 & 0 & 1.5 & -1 & 1 & 0.5 & \tiny $ f_{X_1,X_2}(0,1.5)= f_{\epsilon_1}(-1)   f_{X^*,\epsilon_2}(1,0.5)  $\normalsize \\
3 & 0 & 2 & -1 & 1 & 1 & \tiny $ f_{X_1,X_2}(0,2)= f_{\epsilon_1}(-1)   f_{X^*,\epsilon_2}(1,1)  $\normalsize \\
\hline
4 & -1 & -1.5 & -1 & 0 & -1.5 &  ... \\
5 & -1 & 0.5 & -1 & 0 & 0.5 &  ... \\
6 & -1 & 1 & -1 & 0 & 1 &  ... \\
\hline
7 & 1 & -0.5 & 0 & 1 & -1.5 &  ... \\
8& 1 & 1.5 & 0 & 1 & 0.5 &  ... \\
9 & 1 & 2 & 0 & 1 & 1 &  ... \\
\hline
10 & 0 & -1.5 & 0 & 0 & -1.5 & \tiny $ f_{X_1,X_2}(0,-1,5)= f_{\epsilon_1}(0)  f_{X^*,\epsilon_2}(0,-1.5)  $\normalsize \\
11 & 0 & 0.5 & 0 & 0 & 0.5 &  \tiny $ f_{X_1,X_2}(0,0.5)= f_{\epsilon_1}(0)   f_{X^*,\epsilon_2}(0,0.5)  $\normalsize \\
12 & 0 & 1 & 0 & 0 & 1 &  \tiny $ f_{X_1,X_2}(0,1)= f_{\epsilon_1}(0)  f_{X^*,\epsilon_2}(0,1)  $\normalsize \\
\hline
\end{tabular}
\\ Note: In each group, mean of $X_2$ reveals $X^*$.
\end{center}
\label{table 1b}
\end{table}%

The setup in Equation (\ref{equ 150}) is well known because the distribution of the latent variable $X^{ \ast }$ can be written as a closed-form function of the observed distribution $f_{X_1 ,X_2}$. The characteristic function of $X^{ \ast }$ is defined as $\phi _{X^{ \ast }} (t) =E \left [e^{i t X^{ \ast }}\right ]$ with $i =\sqrt{ -1}$. One can show that 
\begin{eqnarray}f_{X^{ \ast }} \left (x^{ \ast }\right ) &  = & \frac{1}{2 \pi } \int _{ -\infty }^{\infty }e^{ -i x^{ \ast } t} \phi _{X^{ \ast }} \left (t\right )  dt \label{equ 160} \\
\phi _{X^{ \ast }} \left (t\right ) &  = & \exp  \left [\int _{0}^{t}\frac{i E \left [X_2 e^{i s X_1}\right ]}{E \left [e^{i s X_1}\right ]}  ds\right ] . \nonumber \end{eqnarray}  This is the so-called Kotlarski's identity (\cite{kotlarski1966} and \cite{rao_identification}). 
Notice that 
\begin{eqnarray}
\phi _{X_1,X_2}(s,t)  &  = & \exp  \left [ isX_1 + itX_2 \right ] \nonumber  \\
 &  = & \exp  \left [ isX^*  + itX_2 \right ] \exp  \left [  is \epsilon_1 \right ]  \nonumber  \\
  &  = & \phi _{X^*,X_2}(s,t)  \frac{\phi _{X_1}\left (s\right ) }{\phi _{X^{ \ast }} \left (s\right ) } \nonumber 
\end{eqnarray} 
Therefore, we may identify the joint distribution $F_{X^*, X_2}$ as follows:
\begin{eqnarray}
\phi _{X^*,X_2}(s,t)   &  = &  \phi _{X_1,X_2}(s,t)   \frac{\phi _{X^{ \ast }} \left (s\right ) } {\phi _{X_1}\left (s\right ) }\nonumber 
\end{eqnarray} 
Finally, the distribution of $(X_1, X_2, X^*)$ can be uniquely determined by the distribution of $(X_1, X_2)$ as follows:
\begin{eqnarray}
\phi _{X_1, X_2, X^*}(s,t,v)   &  = & \exp  \left [ isX_1 + itX_2 + ivX^* \right ] \nonumber  \\
&  = & \exp  \left [  i(s+v)X^*  + itX_2  \right ]  \exp  \left [  is \epsilon_1 \right ] \nonumber   \\
&  = & \phi _{X^*,X_2}(s+v,t) \frac{\phi _{X_1}\left (s\right ) }{\phi _{X^{ \ast }} \left (s\right ) } 
\end{eqnarray} 

That means $F_{X_1, X_2}$ uniquely determines $F_{X_1, X_2, X^*}$. Under the assumption that observations of $(X_1, X_2)$ are distinct, Condition 1) of Proposition \ref{proposition 1} holds. Then, the identification of $F_{X_1, X_2, X^*}$ implies that the value of $X^*$ can be uniquely determined by that of $(X_1, X_2)$. We summarize the results as follows:

\begin{lemma}
Suppose that $X=(X_1,X_2)$ satisfies Equation (\ref{equ 150}) and that observations of $(X_1, X_2)$ are distinct in population $\mathcal{P}_{X_1, X_2,X^*}$. Then, $\mathcal{P}_{X_1, X_2}$ and $F_{X_1, X_2}$ uniquely determine $\mathcal{P}_{X_1, X_2,X^*}$ and $F_{X_1, X_2, X^*}$, i.e., each $(x_{1,i},x_{2,i})$ in $\mathcal{P}_{X_1, X_2}$ uniquely determines its corresponding $x^*_i$ through $\mathcal{P}_{X_1, X_2,X^*}$.
\end{lemma}

\section{A 3-measurement case}
It is possible to avoid the additivity and linearity in Equation (\ref{equ 150}), when there are more observables. In the case where $X=(X_1,X_2, X_3)$, \cite{Hu2008} provides sufficient conditions to identify the distribution of $(X_1,X_2, X_3, X^*)$ from that of $(X_1,X_2, X_3)$. A version of the conditions is presented here:

\begin{condition}
\label{assumption 2.1}The two measurements $X_1$ and $X_2$ and the latent variable $X^{ \ast }$ share the same support $\mathcal{X} =\left \{v_{1} ,v_{2} ,\ldots  ,v_{K}\right \}$ with $K<N$. 
\end{condition}

\noindent This condition is not restrictive because the results can be straightforwardly extended to the case where supports of measurements $X_1$ and $X_2$ are larger than that of $X^{ \ast }$.

\begin{condition} \label{assumption 2.1b}
The observables satisfy conditional independence as follows:
\begin{equation} \label{cond indep}
F[X_1,X_2,X_3|X^*] = F[X_1|X^*] F[X_2|X^*]  F[X_3|X^*] 
\end{equation}
where $F[X|X^*]$ is the conditional CDF of $X$ on $X^*$.
\end{condition}
Let $f_{X_1 ,X_2}$ be the probability function of $(X_1 ,X_2)$. Define a matrix representation of the joint distribution as follows:
\begin{eqnarray}
M_{X_1 ,X_2 } &  = & \left [f_{X_1 ,X_2} \left (v_{i} ,v_{j} \right )\right ]_{i =1 ,2 ,\ldots  ,K ;j =1 ,2 ,\ldots  ,K} \label{equ 220} 
\end{eqnarray}
We assume

\begin{condition}
\label{assumption 2.2} Matrix $M_{X_1 ,X_2}$ has rank $K$.
\end{condition}

\begin{condition}
\label{assumption 2.3} There exists a function $g  ( \cdot )$ such that $E\left [g  \left (X_3\right )\vert X^{ \ast } =\overline{v}\right ] \neq E\left [g  \left (X_3\right )\vert X^{ \ast } =\widetilde{v} \right ]$ for any $\overline{v} \neq \widetilde{v}$ in $\mathcal{X}$. \end{condition}

\begin{condition}
\label{assumption 2.4} 


$f_{X_1\vert X^{ \ast }} \left (v\vert v\right ) >f_{X_1\vert X^{ \ast }} \left (\widetilde{v} \vert v\right )$ for any $\widetilde{v} \neq v \in \mathcal{X}$, i.e., $v$ is the mode of distribution $f_{X_1|X^*}(\cdot|v)$.

\end{condition}

Under assumptions \ref{assumption 2.1}, \ref{assumption 2.1b}, \ref{assumption 2.2}, \ref{assumption 2.3}, and \ref{assumption 2.4}, \cite{Hu2008} shows that the joint distribution
of the observables $\left (X_1,X_2, X_3\right )$ uniquely determines the joint distribution of the observed variables and the unobservable $\left (X_1,X_2, X_3,  X^*\right )$. Furthermore, if $\left (X_1,X_2, X_3\right )$ satisfies Condition 1) in Proposition  \ref{proposition 1}, then we can apply Proposition  \ref{proposition 1} as follows:

\begin{lemma}  \label{Lemma 2}
Suppose that Assumptions \ref{assumption 2.1}, \ref{assumption 2.1b},  \ref{assumption 2.2}, \ref{assumption 2.3}, and \ref{assumption 2.4} hold, and that observations of $X=(X_1,X_2,X_3)$ are distinct in population $\mathcal{P}_{X_1,X_2, X_3, X^*}$. Then, $\mathcal{P}_{X_1,X_2, X_3}$ and $F_{X_1,X_2, X_3}$ uniquely determine $\mathcal{P}_{X_1,X_2, X_3, X^*}$ and $F_{X_1,X_2, X_3, X^*}$, i.e., each $(x_{1,i},x_{2,i},x_{3,i})$ in $\mathcal{P}_{X_1,X_2, X_3}$ uniquely determines its corresponding $x^*_i$ through $\mathcal{P}_{X_1,X_2, X_3, X^*}$.
\end{lemma}

Because $X_1$ and $X_2$ have a small discrete support, i.e., $K<N$, we need $X_3$ to have a large support to make $X=(X_1,X_2,X_3)$ distinct in the population. For the identification of distributions, $X_3$ can be a measurement as little informative as a binary indictor. But for identification in observations in this paper, $X_3$ needs to have a large support so that $X=(X_1,X_2,X_3)$ is distinct in each observation.


Given the general result above, it is still useful to provide a simple example to illustrate the idea of identification in observations in this case.  Suppose $X_1,X_2,  X^{ \ast }$ share the same support $\{0,1\} $ with non-degenerated misclassification probabilities $f_{X_1|X^*}(1|0) > 0$, $f_{X_1|X^*}(0|1) > 0$, $f_{X_2|X^*}(1|0) > 0$, $f_{X_2|X^*}(0|1) > 0$, and $X_3 \in  \{1,2,3,4\}$  with $f_{X_3|X^*}$ satisfying
\begin{eqnarray*}
f_{X_3|X^*}(1|1) = f_{X_3|X^*}(2|0) = f_{X_3|X^*}(3|0)= f_{X_3|X^*}(4|1) =0.
\end{eqnarray*}
This population is presented in Table \ref{table 2} and satisfies the property of leaves. The goal is to show that the observations of $(X_1,X_2, X_3)$ and the distribution of $(X_1,X_2, X_3)$ can uniquely determine the value of $X^*$ in each observation. 

Assumptions in Lemma  \ref{Lemma 2} hold for the example in Table \ref{table 2}. The conditional independence, together with other assumptions, identifies the probability functions, including  $f_{X_1|X^*}$. Then, for each given value $X_3$, $X^*$ takes a unique value $x^*$, which is equal to the mode of $f_{X_1|X^*}(\cdot|x^*)$ under Assumption \ref{assumption 2.4}. Therefore, the unobserved $x^*$ is uniquely determined by the observed variables in each observation. 

Because every observation of the observables $(X_1,X_2, X_3)$ has to be distinct in the population of $(X_1,X_2, X_3, X^*)$, each value of $(X_1,X_2, X_3)$  can only map to one unique value of $X^*$. It is also useful to present a case where the property of leaves fails. Table \ref{table 3} shows a violation of the property of leaves with $f_{X_3|X^*}$ satisfying
\begin{eqnarray*}
f_{X_3|X^*}(1|1) = f_{X_3|X^*}(2|0) = f_{X_3|X^*}(3|0) =0 \\
f_{X_3|X^*}(4|0) >0 \, ,  \, f_{X_3|X^*}(4|1) >0 
\end{eqnarray*}
in the example above because observations 13, 14, 15, 16 are the same as observations 17, 18, 19, 20, respectively, if we only observe $(X_1,X_2, X_3)$. In other words, $X_3=4$ corresponds to $X^*=0$ and $X^*=1$. In particular, leaves (or observations) 13 and 17 are different in population, but they are the same from a researcher's view because they don't observe $X^*$. That is the case we rule out here because it is not consistent with the common knowledge that no two leaves are alike.

\begin{table}[htp]
\caption{An illustration of identification in observations}
\begin{center}
\begin{tabular}{c|ccc|c|c}
\hline 
\hline
observation & \multicolumn{3}{c|}{observables} &\multicolumn{1}{c|}{unobservables}& probability\\
$i$ & $X_1$ & $X_2$  & $X_3$   &  $X^{ \ast }$ &  $p_i $\\
\hline
1 & 0 & 0 & 1  & 0 & \tiny $  f_{X_1,X_2,X_3}(0,0,1) = f_{X_1|X^*}(0|0)f_{X_2|X^*}(0|0)f_{X_3|X^*}(1|0) f_{X^*}(0)$ \normalsize \\
2 & 1 & 0 & 1  & 0 &  \tiny $ f_{X_1,X_2,X_3}(1,0,1) = f_{X_1|X^*}(1|0)f_{X_2|X^*}(0|0)f_{X_3|X^*}(1|0) f_{X^*}(0)  $ \normalsize \\
3 & 0 & 1 & 1  & 0 & ... \\
4 & 1 & 1 & 1  & 0 & ... \\
\hline
5 & 0 & 0 & 2  & 1 & ... \\
6 & 1 & 0 & 2  & 1 & ... \\
7 & 0 & 1 & 2  & 1 & ... \\
8 & 1 & 1 & 2  & 1 & ... \\
\hline
9 & 0 & 0 & 3  & 1 & ... \\
10 & 1 & 0 & 3  & 1 & ... \\
11 & 0 & 1 & 3  & 1 & ... \\
12 & 1 & 1 & 3  & 1 & ... \\
\hline
13 & 0 & 0 & 4  & 0 & ... \\
14 & 1 & 0 & 4  & 0 & ... \\
15 & 0 & 1 & 4  & 0 & ... \\
16 & 1 & 1 & 4  & 0 & ... \\
\hline

\end{tabular}
\\ Note: For a given $X_3$, $X^*$ takes a unique value  equal to the mode of $f_{X_1|X^*}(\cdot|x^*)$.\\
\end{center}
\label{table 2}
\end{table}%

\begin{table}[htp]
\caption{A violation of the property of leaves in Equation  \ref{definition 1}}
\begin{center}
\begin{tabular}{c|ccc|c|c}
\hline 
\hline
observation & \multicolumn{3}{c|}{observables} &\multicolumn{1}{c|}{unobservables}& probability\\
$i$  & $X_1$ & $X_2$  & $X_3$   &  $X^{ \ast }$ &  $p_i $\\
\hline
1 & 0 & 0 & 1  & 0 & \tiny $  f_{X_1,X_2,X_3}(0,0,1) = f_{X_1|X^*}(0|0)f_{X_2|X^*}(0|0)f_{X_3|X^*}(1|0) f_{X^*}(0)$ \normalsize \\
2 & 1 & 0 & 1  & 0 &  \tiny $ f_{X_1,X_2,X_3}(1,0,1) = f_{X_1|X^*}(1|0)f_{X_2|X^*}(0|0)f_{X_3|X^*}(1|0) f_{X^*}(0)  $ \normalsize \\
3 & 0 & 1 & 1  & 0 & ... \\
4 & 1 & 1 & 1  & 0 & ... \\
\hline
5 & 0 & 0 & 2  & 1 & ... \\
6 & 1 & 0 & 2  & 1 & ... \\
7 & 0 & 1 & 2  & 1 & ... \\
8 & 1 & 1 & 2  & 1 & ... \\
\hline
9 & 0 & 0 & 3  & 1 & ... \\
10 & 1 & 0 & 3  & 1 & ... \\
11 & 0 & 1 & 3  & 1 & ... \\
12 & 1 & 1 & 3  & 1 & ... \\
\hline
13 & 0 & 0 & 4  & 0 & \tiny $   f_{X_1|X^*}(0|0)f_{X_2|X^*}(0|0)f_{X_3|X^*}(4|0) f_{X^*}(0)$ \\
14 & 1 & 0 & 4  & 0 & ... \\
15 & 0 & 1 & 4  & 0 & ... \\
16 & 1 & 1 & 4  & 0 & ... \\
\hline
17 & 0 & 0 & 4  & 1 & \tiny $   f_{X_1|X^*}(0|1)f_{X_2|X^*}(0|1)f_{X_3|X^*}(4|1) f_{X^*}(1)$ \\
18 & 1 & 0 & 4  & 1 & ... \\
19 & 0 & 1 & 4  & 1 & ... \\
20 & 1 & 1 & 4  & 1 & ... \\
\hline
\end{tabular}
\\ Note: For $X_3=4$, $X^*$ is not unique.\\
 $ f_{X_1,X_2,X_3}(0,0,4) = \sum_{x^* \in \{0,1\}} f_{X_1|X^*}(0|x^*)f_{X_2|X^*}(0|x^*)f_{X_3|X^*}(4|x^*) f_{X^*}(x^*)$
\end{center}\label{table 3}
\end{table}%

\section{Summary}

This paper provides sufficient conditions for the identification of unobserved variables in observations at the population level. Based on an observed feature of the data -- the property of leaves, the results in this paper imply that when the joint distribution of the observable $X$ and the unobservable $X^*$ satisfies certain conditions, it is not only possible to identify their joint distribution $F_{X, X^* }$ from $F_{X }$, but also possible to identify the value of the unobservable $X^*$ in observations.  The distinctness of observed variables in observations implies there exists a function mapping from the observables to the unobservables. Such a function guarantees the uniqueness of the latent value in each observation. The joint distribution can then reveal the latent value in each observation.

In the identification analysis, we consider distribution function $F_{X }$ to be identified as the limit of the empirical distribution of $X$ from a sample. The results in this paper suggests that it is possible to use the sample counterpart of this argument to estimate unobservables at the observation level. A simple existing example is OLS residuals in a linear regression model. The identification results here imply that researchers may tackle unobservables by directly estimating them in a broad range of models. 


\bibliographystyle{aer}
\bibliography{Ref}
\end{document}